\documentclass[a4paper]{jpconf}
\usepackage{graphicx}

\usepackage{amsmath}
\usepackage[utf8]{inputenc}
\usepackage[]{lineno}
\usepackage{subfigure}
\newcommand{\GERDA}{\textsc{Gerda}}

\newcommand{\GELATIO}{\textsc{Gelatio}}

\begin{document}
\title{Off-line data quality monitoring for the GERDA experiment}

\author{P Zavarise$^{1,2}$, M Agostini$^3$, A A Machado$^{2,4}$, 
L Pandola$^2$ and O~Volynets$^5$}
\address{$^1$ Universit\`a dell'Aquila, L'Aquila, Italy}
\address{$^2$ INFN, Laboratori Nazionali del Gran Sasso, Assergi, Italy}
\address{$^3$ Physik Department E15, Technische Universit\"at M\"unchen, Garching, Germany}
\address{$^4$ Max-Planck-Institut f\"ur Kernphysik, Heidelberg, Germany}
\address{$^5$ Max-Planck-Institut f\"ur Physik, Munich, Germany}
\ead{zavarise@lngs.infn.it}
\begin{abstract}
\GERDA\ is an experiment searching for the neutrinoless $\beta\beta$ decay
of $^{76}$Ge. The experiment uses an array of high-purity germanium detectors, 
enriched in $^{76}$Ge, directly immersed in liquid argon. 
\GERDA\ recently started the physics data taking using eight enriched coaxial 
detectors. 
The status of the experiment has to be closely monitored in order to promptly 
identify possible instabilities or problems. 
The on-line slow control system is 
complemented by a regular off-line monitoring of data quality.
This ensures that data are qualified to be used in the physics 
analysis and allows to reject data sets which do not meet the minimum quality 
standards.
The off-line data monitoring is entirely performed within the software 
framework \GELATIO.
In addition, a relational database, 
complemented by a web-based interface, was developed to support the off-line 
monitoring and to automatically provide information to daily assess 
data quality. The concept and the performance of the off-line monitoring 
tools were tested and validated during the one-year commissioning phase.
\end{abstract}

\section{Introduction}
\GERDA\ is a low-background experiment which searches for the neutrinoless 
$\beta\beta$ decay of $^{76}$Ge, using an array of high-purity germanium (HPGe) 
detectors isotopically enriched in $^{76}$Ge~\cite{gerda}.
The detectors are operated naked in ultra radio-pure liquid argon, which acts 
as the cooling medium and as a passive shielding against the external $\gamma$ 
radiation. This innovative design $-$ complemented by a strict material selection for 
radio-purity $-$ allows to achieve an unprecedented background level in the region 
of the $Q_{\beta\beta}$-value of $^{76}$Ge at 2039~keV. 
The experiment is located in the underground Laboratori Nazionali del Gran Sasso 
of the INFN (Italy).
The Phase~I of \GERDA\ recently started using eight enriched coaxial detectors 
(totaling approximately 15~kg of $^{76}$Ge). The Phase~I comes after a one-year-long 
commissioning, in which natural and enriched HPGe detectors were 
successfully operated in the \GERDA\ set-up.\\
\GERDA\ has an advanced slow control system to monitor the main parameters characterizing the operational status of 
the sub-systems, as temperature, pressure, detector currents, and to handle alarms. 
A different kind of monitoring $-$ the ``data quality'' monitoring $-$ is also 
crucial to check the overall performance of the experiment and to certify the validity 
of the data for physics analysis. A key difference between the slow control system 
and the data quality monitoring is that the latter is not performed in real-time, but 
relies on an off-line data analysis.
Both approaches are needed to ensure that every sub-system in \GERDA\ is running 
in the intended way and that the data produced by the experiment are qualified for 
physics analysis.
\section{The GERDA data flow}
The data collected by the main \GERDA\ acquisition system (DAQ) consist of charge signals
from the HPGe detectors, digitized at 100~MHz sampling rate. Raw data files 
are automatically copied once per day, at night-time, from the DAQ server located in 
the underground laboratory to the main \GERDA\ file server, which is located in the 
above-ground facilities of the Gran Sasso Laboratory. New files are automatically processed 
using the framework \GELATIO, 
which was developed for the \GERDA\ experiment~\cite{gelatio}. Data 
are analyzed within \GELATIO\ with a modular approach, according to the 
procedure described in Ref.~\cite{acat}. The analysis modules implemented in 
\GELATIO\ are tailored to extract physics information from the signals, like rise 
time and amplitude. The parameters produced as output by \GELATIO\ 
are also imported in a relational database system, based on 
MySQL \cite{mysql}. 
The database is fully integrated with \GELATIO, in the 
sense that it allows for a two-way relationship, i.e. in both input and output. 
For instance it is possible to create a \GELATIO-compliant file from a 
selection of events made by the database using SQL.
An advanced web interface was developed to allow an easy and 
user-friendly interaction with the database. 
It can be used to send queries and to generate a set of 
standardized reports, which contain diagnostic information 
to assess the data quality.
Reports are issued in the form either of numbers (e.g. counting rates) or of graphs
(1-dimensional histograms, scatter plots).
More details are presented in the following sections.
\section{Duty cycle monitoring}
A key diagnostic parameter is the duty cycle of the experiment within a 
given set of ``Runs''. A ``Run'' is defined by the geometrical configuration (number and position 
of the HPGe detectors) and the bias 
high voltage applied to the detector electrodes.
If everything is working properly, data acquisition during a Run is interrupted only when 
a calibration of the system has to be performed. A calibration consists in the irradiation 
of the detectors by $^{228}$Th radioactive sources. It is typically performed once per week 
and the procedure requires less than two hours in total. Being \GERDA\ a 
low-rate experiment, the amount of data is largely dominated by calibrations. 
Interruptions of the normal data taking may happen because either  
of scheduled operations (maintenance, upgrade, ancillary measurements) or of 
hardware problems (e.g. electrical power spikes or instabilities, failure of a sub-system). 
The overall duty cycle achieved during the last six months of the \GERDA\ commissioning, 
excluding the interruptions due to system upgrades, is larger than 90\%, 
consistently with the expectations. Given the low counting rate, 
no loss in duty cycle is expected from the dead time of electronics. 
\section{Rate monitoring} \label{sec:rates}
\begin{figure}[t]
\[
\begin{smallmatrix}
\includegraphics[width=7.8cm]{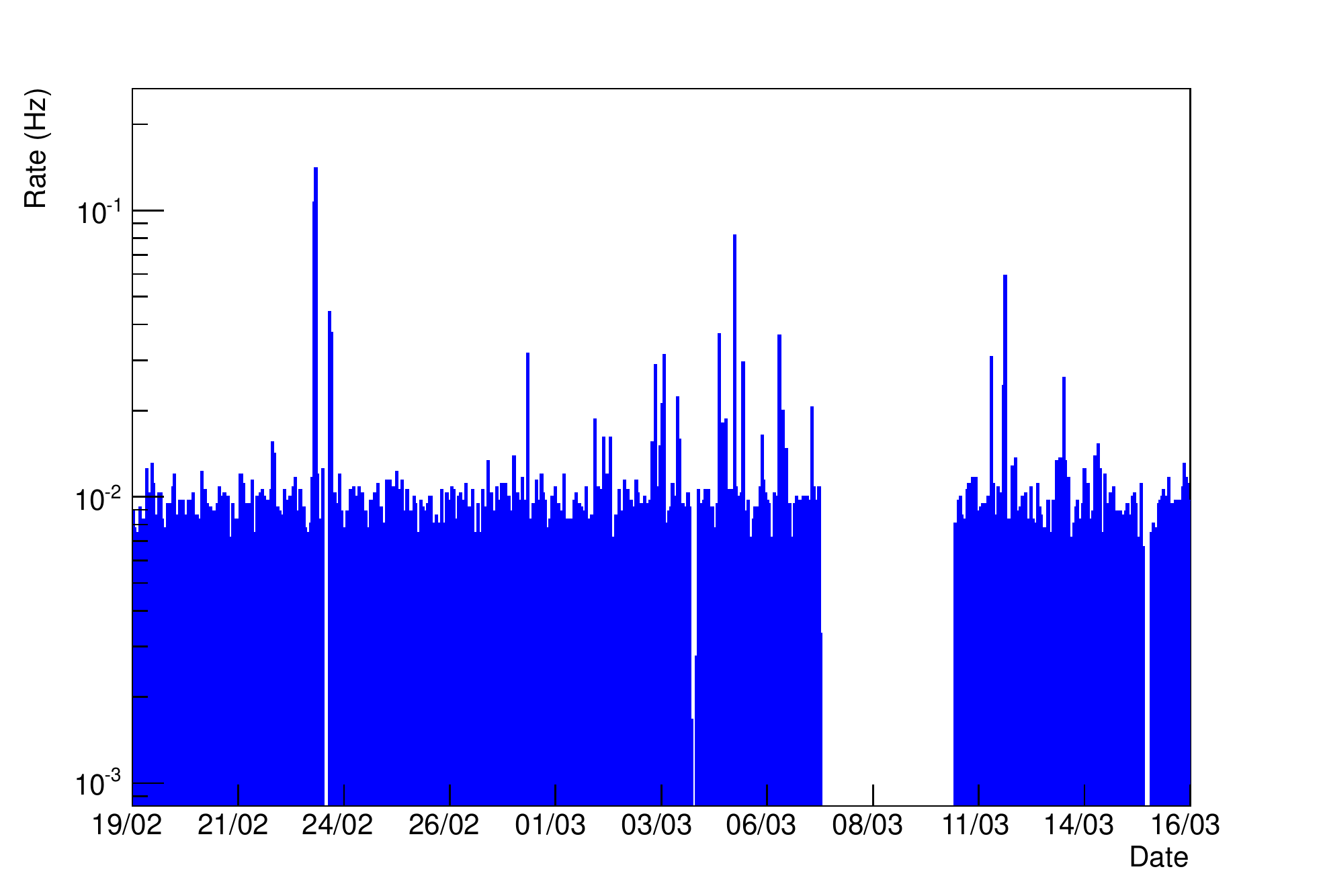}&\hspace{0.5cm}&
\includegraphics[width=7.8cm]{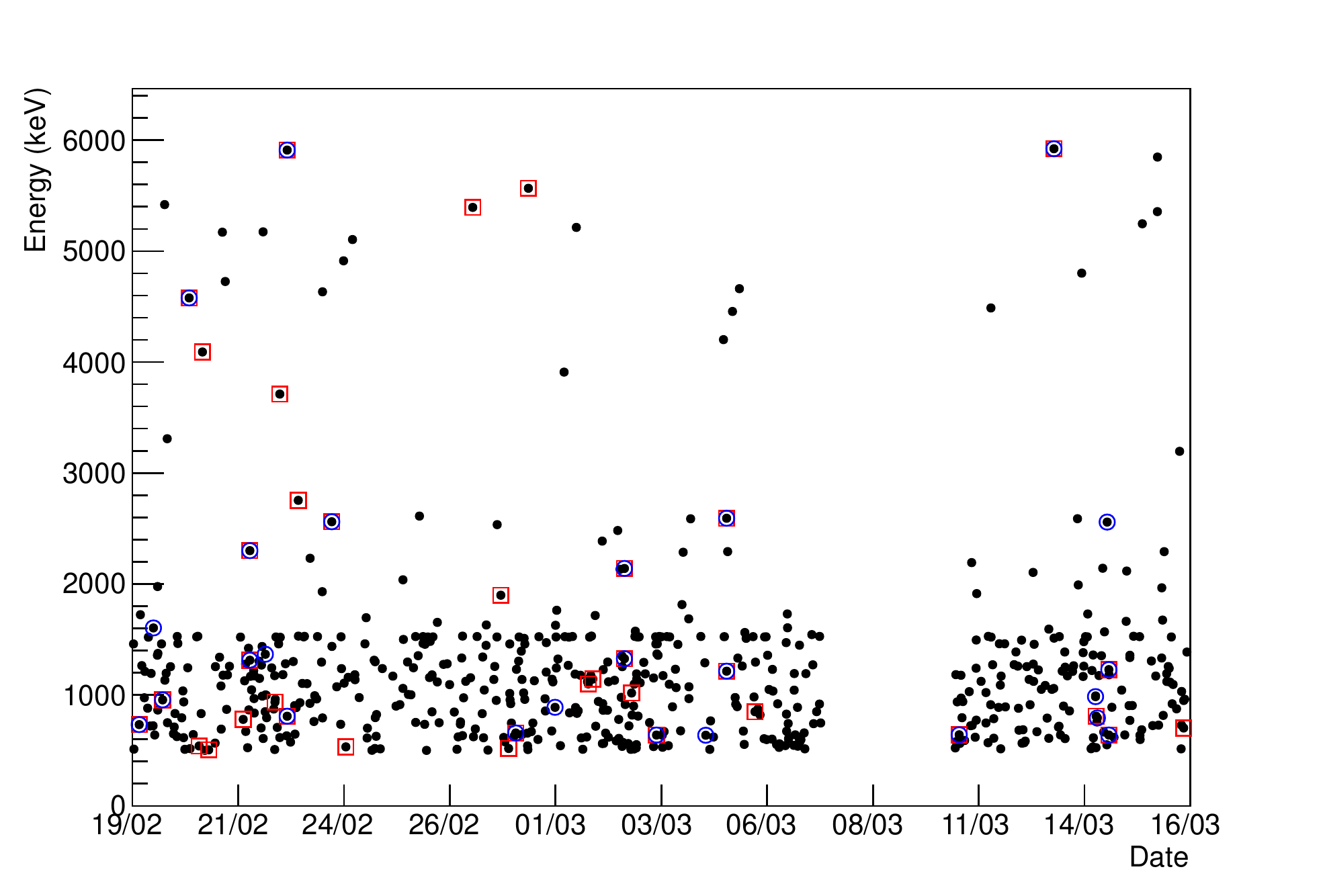} \\
\text{(a) Time distribution}&&\text{(b) Energy vs. time distribution}
\end{smallmatrix}
\]
\caption{\small Illustrative plots from the database daily report. 
(a) Counting rate of physical events vs. time. A few noise bursts are clearly visible, 
in which the counting rate is increased by a factor $> 2$. (b) Energy vs. time scatter plot for 
physical events above 500~keV. Events with more than one HPGe detector fired are marked by 
a red square, while events that are in coincidence with the muon veto system are marked 
by a blue circle. The 3-day empty region is due to an instability of the electronic system, which 
caused the data to fail the quality checks. The plots refer to an early \GERDA\ commissioning Run 
performed on February-March 2011 with three non-enriched detectors.}
\label{fig:timedistplots}
\end{figure}

The counting rates for different classes of events, as physical events 
in the HPGe detectors and ``test pulses'', are also important parameters to 
be monitored.
During the data taking a pulse of fixed amplitude (test pulse) is automatically 
produced by a pulser and fed to all DAQ channels, in order 
to monitor the stability of the electronics. Test pulses 
are issued at the rate of 0.1~Hz, which is much higher than 
the typical rate of physical events observed in \GERDA\ (about 2~mHz per detector in 
the full energy spectrum\footnote{The energy threshold of the HPGe detectors
during the commissioning was typically between 50 and 100~keV.}). 
More than 80\% of the events collected in an ordinary \GERDA\ run are test pulses. \\
The total event rate in the HPGe detectors is expected to be 
approximately constant in time. The possible occurrence of noise bursts which trigger the DAQ system 
can cause a substantial increase of the event rate with respect to the normal value. 
The characteristic signature of noise bursts in the event rate distribution can be observed  
in the Fig.~\ref{fig:timedistplots}a. \\
In order to tag and reject events in the HPGe detectors due to muon-induced 
interactions \GERDA\ features a muon veto system~\cite{muon}. 
The general trigger of the muon veto system (which is a 1~$\mu$s-long NIM logical signal) 
is fed to the data acquisition of the HPGe detectors. For each interaction in the germanium  
array, the logical muon veto signal is hence encoded in the data stream. This permits 
the evaluation of the rate of muon-induced events in the HPGe detectors: the typical 
rate measured in the most recent Runs is about 0.5~counts/day per detector. 
It is in agreement with the expectations based on the 
Monte Carlo simulation performed with the framework \textsc{MaGe}~\cite{mage} 
according to the procedure of Ref.~\cite{muonMC}.
Figure~\ref{fig:timedistplots}b shows the energy vs. time scatter plot for physical events 
detected in an early \GERDA\ commissioning run with three non-enriched detectors. The occurrence 
of noise bursts and of interruptions due to poor data quality was significantly reduced 
in the most recent commissioning runs.
\section{Monitoring of the read-out electronics performance}
\begin{figure}[t]
\[
\begin{smallmatrix}
\includegraphics[width=7.8cm]{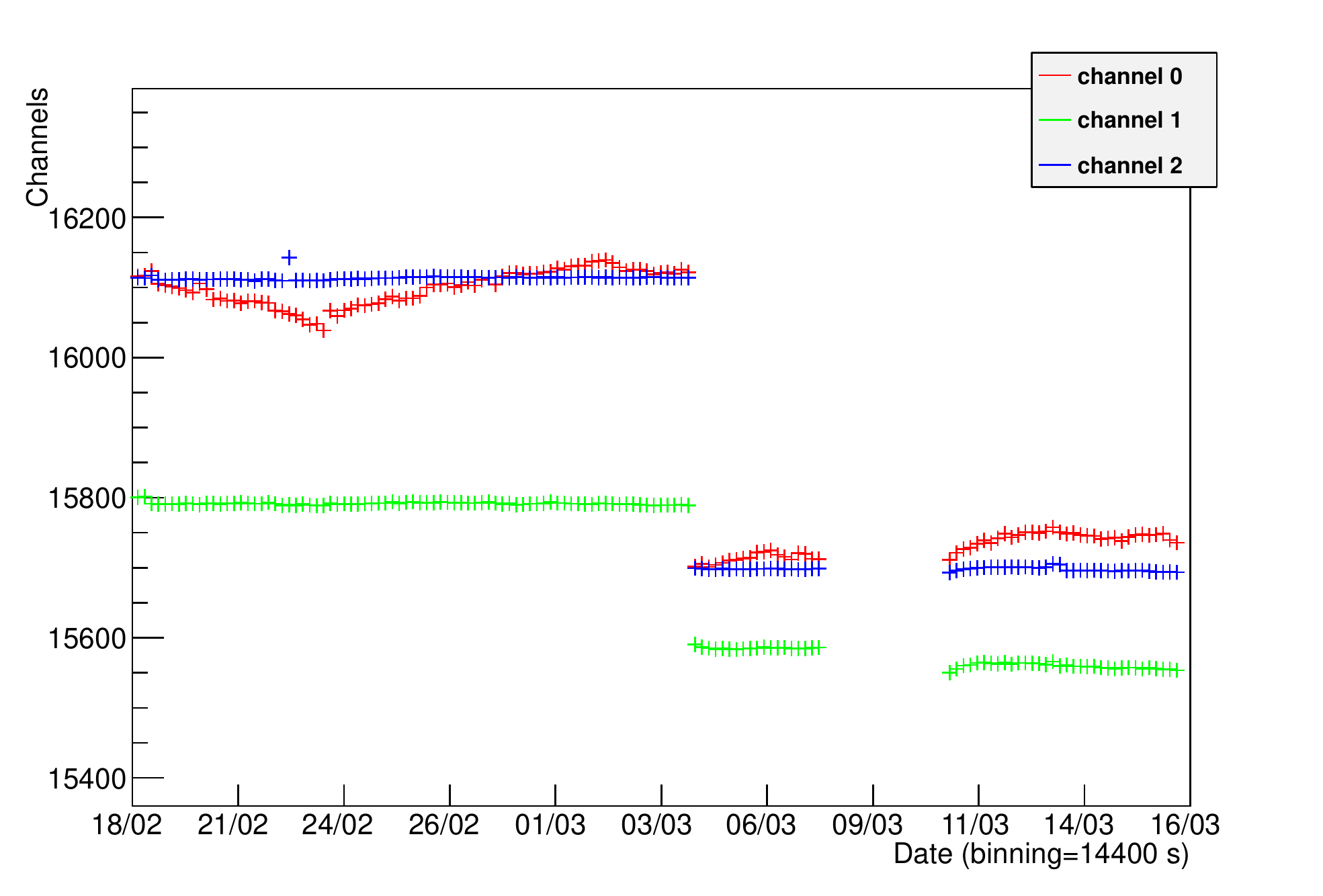}&\hspace{0.5cm}&
\includegraphics[width=7.8cm]{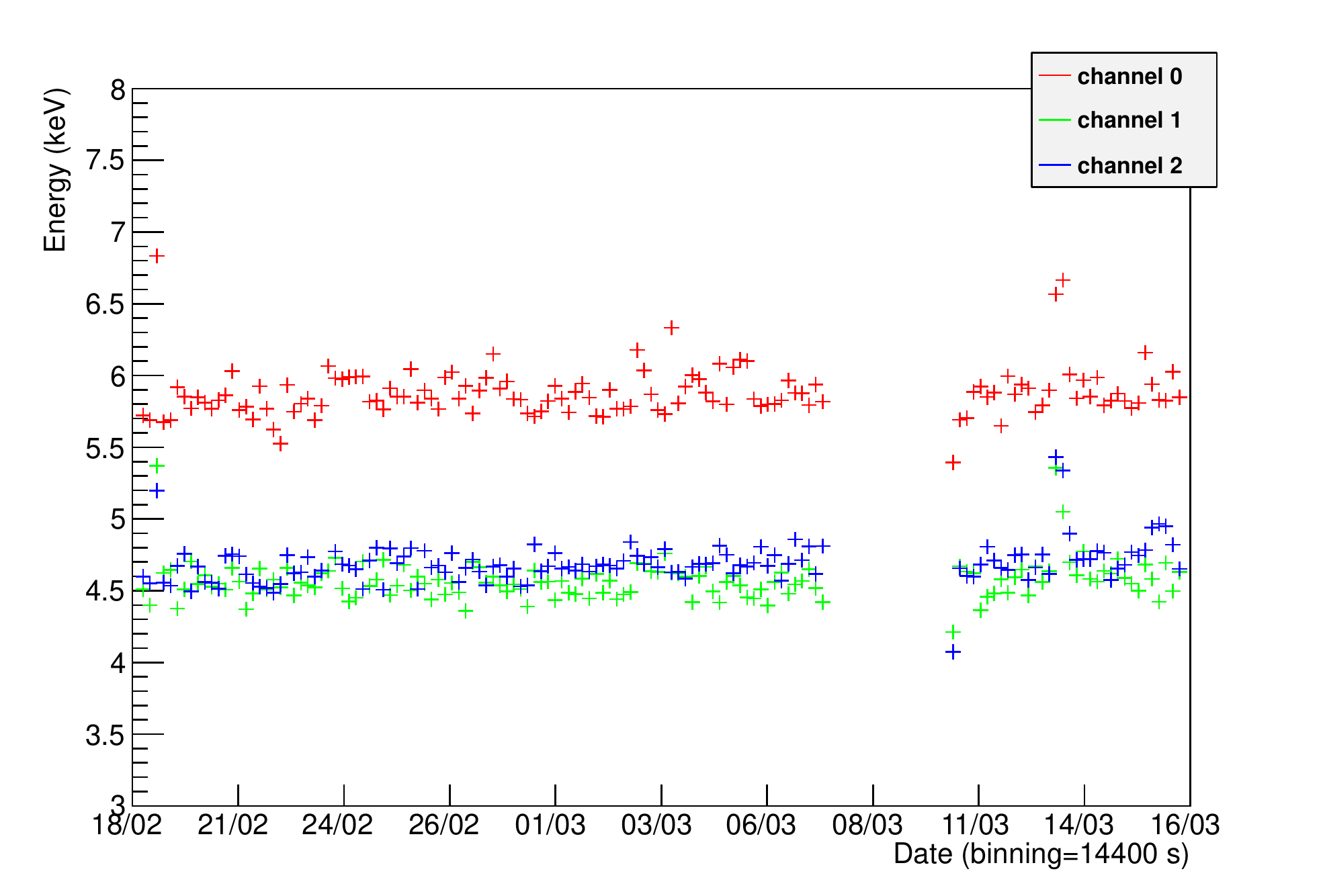}\\
\text{(a) Baseline amplitude}&&\text{(b) Test pulse equivalent energy resolution }\\
\end{smallmatrix}
\]
\caption{\small Monitoring of the stability in time of the electronic chain. (a) Baseline amplitude vs time. (b) Energy resolution of the test pulse line (FWHM) vs. time.}
\label{fig:pulser}
\end{figure}

To monitor the read-out electronic chain and DAQ system stability, the database analyzes several parameters
including:
\begin{enumerate}
\item Amplitude of the baseline vs. time (Fig.~\ref{fig:pulser}a). 
Fluctuations or drifts in the position of the baseline may indicate changes in the leakage 
current of the HPGe detectors or in the gain of the electronic chain.
\item Root-mean-square (rms) of the baseline vs. time. The fluctuations of 
the baseline position with respect to the average value are a direct measurement of the noise 
of the electronic chain. Variations or sudden shifts of the baseline rms are symptoms of changes 
in the operation of the electronic chain. 
\item Test pulse amplitude vs. time. 
Since the signal injected in the electronic 
chain is constant, a time variation of its amplitude indicates a change in the 
global response of the electronic chain, e.g. gain drift, change in the system capacitance, etc.
\item Width of the test pulse line vs. time (Fig.~\ref{fig:pulser}b). Fluctuations of the width 
of test pulse line are related to the electronic noise of the chain. They are quoted 
as full width at half maximum (FWHM) of the peak, as customary in $\gamma$-ray spectroscopy.
\end{enumerate}
When an anomalous behavior is observed in one of these indicators,
the corresponding data are marked as invalid and not included in the 
reference set for the subsequent analysis.
%
%
\section{Conclusions}
A complete system has been developed  
for the off-line monitoring of the data quality of the \GERDA\ experiment. 
A standard set of daily reports and graphs to assess the stability of the system, which 
complements the information coming from the slow control, can be obtained by the combined use of \GELATIO\ 
and of the dedicated \GERDA\ database system. The concept and the performance of the off-line monitoring 
tools have been tested and validated during the one-year commissioning phase. All tools were found to be 
properly working and will be routinely used in the Phase~I of the experiment.
\section*{Acknowledgments}
We wish to thank our Colleagues of the \GERDA\ Collaboration, especially C.~Cattadori, for their useful 
feedback and suggestions. 
We also express our gratitude to S.~Stalio - from the LNGS IT Service - for the continuous 
support in many technical issues related to the \GERDA\ main server. \\

\end{document}